\documentstyle[11pt,epsf]{article}
\newcommand{\postscript}[2]
   {\setlength{\epsfxsize}{#2\hsize}
   \centerline{\epsfbox{#1}}}
\begin{document}
\def\theequation {\thesection.\arabic{equation}}
\makeatletter\@addtoreset {equation}{section}\makeatother
\title{\bf A tail-matching method for the linear stability of multi-vector-soliton bound states}
\author{Jianke Yang\thanks{This work was supported in part by a NASA EPSCoR grant. }}
\date{\empty}
\maketitle
\thispagestyle{empty}

\begin{abstract}
Linear stability of multi-vector-soliton bound states in the coupled nonlinear 
Schr\"odinger equations is analyzed using a new tail-matching method. 
Under the condition that individual vector solitons in the bound states
are wave-and-daughter-waves and widely separated, small eigenvalues of these bound states
that bifurcate from the zero eigenvalue of single vector solitons
are calculated explicitly.   
It is found that unstable eigenvalues from phase-mode bifurcations
always exist, thus the bound states are always linearly unstable. 
This tail-matching method is simple, but it gives identical results as 
the Karpman-Solev'ev-Gorshkov-Ostrovsky method. 

\end{abstract}

{\em Mathematics Subject Classification:} 35Q55, 35Pxx, 74J35. 

{\bf Key words: } linear stability; coupled NLS equations; tail-matching method. 

\section{Introduction}
Nonlinear optics and fiber communication systems are advancing very rapidly
these days. In this process, a widely-used mathematical model is the
coupled nonlinear Schr\"odinger (NLS) equations which govern 
pulse propagation in birefringent fibers \cite{menyuk87,agrawal,hasegawa}. 
Similar equations with a saturable nonlinearity also govern 
the interaction of two incoherently-coupled laser beams \cite{anastassiou,kivshar}. 
These equations arise in water-wave interactions as well \cite{benney,roskes}. 
Solution properties of the coupled NLS equations have been examined 
extensively in the past ten years. It is known that
these equations admit single-hump vector solitons due to a nonlinear mutual
trapping effect \cite{menyuk88,YangPhysica97}. When these solitons are perturbed, 
they undergo long-lived internal oscillations 
\cite{ueda_kath,malomedkaup,YangStudies97}. 
When they collide with each other, a fractal structure can arise in the parameter 
space \cite{YangTanPRL,TanYangPRE}. 

Multi-vector-soliton bound states also exist in the coupled NLS equations
\cite{Haelterman93}. These states are pieced together by several
single-hump vector solitons. They received much attention because
of several reasons. First, in fiber communication systems, pulse-pulse 
interference impairs the system performance. If multi-soliton bound states
exist, these solutions would have implications to system designs. 
Second, the existence of such bound states is noteworthy because
they can not exist in the NLS equation \cite{zakharov}. 
Thirdly, these states are closely related to similar states in 
incoherent laser beams, which have been observed experimentally \cite{chen00}. 

After the numerical discovery of these multi-vector-soliton bound states in 
\cite{Haelterman93}, their analytical construction was made 
in \cite{YangStudies98} by a tail-matching technique. 
Their linear-stability problem was studied later in \cite{YangPRE01} by an 
extension of the Karpman-Solev'ev-Gorshkov-Ostrovsky (KSGO) method 
\cite{karpman,gorshkov}, 
and small eigenvalues bifurcated from the zero eigenvalue 
of single vector solitons were calculated. These calculations show that
multi-soliton bound states are always linearly unstable. 
But this KSGO method is quite involved, thus a simpler technique for the calculation of 
eigenvalue bifurcations is called upon. 

In this paper, we use a new tail-matching method to analyze the linear stability 
of two-vector-soliton bound states in the coupled NLS equations. Under the condition that 
individual vector solitons in these bound states are wave-and-daughter-waves 
(i.e., one component is much larger than the other one), and are 
widely separated, small eigenvalues of these bound states 
that bifurcate from the zero eigenvalue of single solitons are calculated.
These small eigenvalues are all the non-zero discrete eigenvalues of the two-soliton bound states.  
We found that unstable eigenvalues from phase-mode bifurcations always exist, thus 
the bound states are always linearly unstable.  
The present technique is much simpler, but it gives identical results as
the KSGO method \cite{YangPRE01}. 

\section{Two-vector-soliton bound states: a review}
The coupled NLS equations
\begin{equation}\label{A}
iA_t+A_{xx}+(|A|^2+\beta|B|^2)A=0, 
\end{equation}
\begin{equation}\label{B}
iB_t+B_{xx}+(|B|^2+\beta|A|^2)B=0, 
\end{equation}
govern optical pulse propagation in birefringent fibers \cite{menyuk87,agrawal,hasegawa}. 
Here $\beta$ is the cross-phase-modulational coefficient. When $\beta=0$ or 1, 
these equations are integrable by the inverse scattering method \cite{zakharov,manakov}. 

These equations admit solitary-wave solutions of the following form: 
\begin{equation} \label{vectorsoliton}
A(x, t)=r(x) e^{i\omega^2 t}, \hspace{0.4cm} B(x, t)=R(x) e^{i\Omega^2 t},
\end{equation}
where $\omega$ and $\Omega$ are frequencies, and the real-valued amplitude functions 
$r(x)$ and $R(x)$ satisfy the ordinary differential equations (ODEs): 
\begin{equation} \label{r}
r_{xx}-\omega^2 r + (r^2+\beta R^2) r=0, 
\end{equation}
\begin{equation} \label{R}
R_{xx}-\Omega^2 R + (R^2+\beta r^2) R=0. 
\end{equation}
After a simple rescaling of variables, we normalize $\omega=1$. 
Since Eqs. (\ref{A}) and (\ref{B}) are Galilean-invariant, 
moving solitary waves can be readily deduced from the stationary ones
(\ref{vectorsoliton}) (see \cite{yangbenney}). 

Solitary waves in Eqs. (\ref{r}) and (\ref{R}) have been studied extensively
before (see \cite{YangPhysica97} and the references therein). 
It has been shown that for any frequency $\Omega \in [\Omega_c, 1/\Omega_c]$
where 
\begin{equation}\label{Omegac}
\Omega_c\equiv \frac{\sqrt{1+8\beta}-1}{2},
\end{equation}
this ODE system admits a unique, single-hump, and positive vector-soliton solution
which is symmetric in both $r$ and $R$ components. We call this solution
the fundamental vector soliton, and denote it as $[r_0(x), R_0(x)]$. 
The asymptotic behavior of this fundamental soliton at infinity is
\begin{equation} \label{largex}
r_0(x) \longrightarrow c\: e^{-|x|}, \hspace{0.5cm}
R_0(x) \longrightarrow C e^{-\Omega |x|}, \hspace{0.8cm} |x| \rightarrow \infty, 
\end{equation}
where $c$ and $C$ are tail coefficients. 
When $\Omega$ is close to the boundary value $\Omega_c$, $R_0(x) \ll r_0(x)$,
$c\approx 2\sqrt{2}$, and $C\ll 1$; 
if $\Omega$ is close to $1/\Omega_c$, $r_0(x)\ll R_0(x)$, 
$c\ll 1$, and $C\approx 2\sqrt{2}/\Omega_c$. These vector solitons
with $R_0\ll r_0$ or $r_0 \ll R_0$ are the so-called wave-and-daughter-waves. 

\begin{figure}[h]
\begin{center}
\parbox[t]{12cm}{\postscript{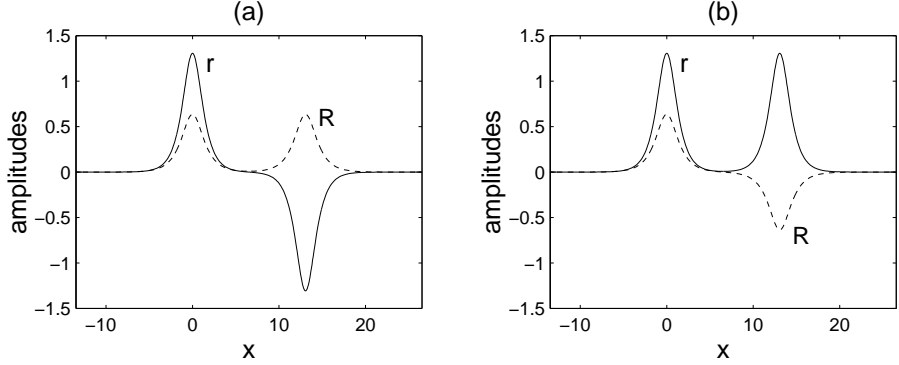}{1.0}}

\vspace{0.5cm}
\caption{Numerically obtained 
stationary two-vector-soliton bound states at $\beta=2/3$ and
$\Omega=0.85$: (a) type-I state; (b) type-II state. 
Analytical approximations by Eqs. (\ref{rsum}), (\ref{Rsum}), and 
(\ref{spacing}) are indistinguishable 
from the numerical curves and thus not shown.
\label{2wave} }  
\end{center} 
\end{figure}

Two-vector-soliton bound states also exist in the ODE system (\ref{r}) and (\ref{R})
\cite{YangPhysica97,Haelterman93,YangStudies98,YangPRE01}. These bound states
look like two single-humped vector solitons glued together, while the two solitons 
are in-phase in one component, and out-of-phase in the other component. 
The in-phase component of the bound states are symmetric around the bound-state center, 
and the out-of-phase component are anti-symmetric around the bound-state center. 
In the limit of large soliton separation, these bound
states approach a superposition of two fundamental solitons (to the leading order): 
\begin{equation} \label{rsum}
r(x)\longrightarrow r_0(x)\mp r_0(x-\Delta), 
\end{equation}
\begin{equation} \label{Rsum}
R(x)\longrightarrow R_0(x)\pm R_0(x-\Delta), 
\end{equation}
where the separation $\Delta \gg 1$. 
These widely-separated 
bound states exist in two parameter-regions \cite{YangPhysica97,Yangunpublished}:  
(i) $\Omega \approx \Omega_c$ or $1/\Omega_c$, and $0<\beta<1$; 
(ii) $\Omega \approx 1$, and $\beta>0$. 
In the first region, the bound states look like two wave-and-daughter-waves glued together; 
while in the second region, the bound states look like two nearly-equal-amplitude 
vector solitons glued together. 
In this article, we only consider the bound states in the first region. 
In this region, the spacing between the two wave-and-daughter-waves in the bound state
is given by the formula \cite{YangStudies98,YangPRE01}
\begin{equation} \label{spacing}
\Delta =\frac{\ln c^2-\ln \Omega^2C^2}{1-\Omega}.
\end{equation}
Below, the bound states with the minus sign in (\ref{rsum}) and the plus sign
in (\ref{Rsum}) will be termed type-I bound states, while the ones with
the plus sign in (\ref{rsum}) and the minus sign in (\ref{Rsum}) termed 
type-II bound states (as we have done in \cite{YangPRE01}). 
These bound states at parameter values $\beta=2/3$ and $\Omega=0.85$ are displayed in Fig. \ref{2wave}. 
A comparison between the analytical spacing formula (\ref{spacing}) and 
numerical values has been made in \cite{YangPRE01}, and excellent agreement
has been obtained. 

Analytical construction of multi-soliton bound states in a general nonlinear wave system was made 
in \cite{YangStudies98} using a tail-matching method, and the spacing formula (\ref{spacing})
for the coupled NLS equations was derived only as a special case
(see \cite{ChampneysYang} for an application of this method for
the construction of other types of multi-pulse bound states).
Below, we use a simplified version of \cite{YangStudies98}'s method to construct
multi-vector-soliton bound states in the  
coupled NLS equations [i.e., (\ref{r}) and (\ref{R})], and reproduce 
the spacing formula (\ref{spacing}). 
There are two reasons for our doing this: (i)
to highlight the key ideas in the tail-matching method for the construction of multi-pulse bound states; 
(ii) to motivate a similar tail-matching idea for the linear-stability analysis
of multi-pulse bound states (see Sec. 3). 

Suppose we have a bound state of two vector solitons located at $x=0$ and $\Delta \:(\gg 1)$.  
As $\Delta \to \infty$, the leading-order asymptotics of this bound state is
\begin{equation} \label{rRlimit}
r(x) \longrightarrow r_0(x)+s_1 r_0(x-\Delta), \hspace{0.5cm}
R(x) \longrightarrow R_0(x)+s_2 R_0(x-\Delta),
\end{equation}
where $s_1$ and $s_2$ are sign-constants and are either 1 or $-1$. 
Our task is to determine values of $s_1, s_2$ and the spacing $\Delta$. 
For this purpose, we consider the bound state in two $x$-regions: 
$-\infty<x\ll \Delta$, and $0\ll x<\infty$. Since the treatments for these two regions are
the same, we only look at the first region $-\infty<x\ll \Delta$. In this region, 
the bound state is a slightly perturbed fundamental vector soliton, i.e., 
\begin{equation} \label{rRperturb}
r(x)=r_0(x)+\tilde{r}(x), \hspace{0.5cm} R(x)=R_0(x)+\tilde{R}(x), 
\end{equation}
where $\tilde{r}, \tilde{R} \ll 1$. 
The actual forms and sizes of $\tilde{r}$ and $\tilde{R}$ are not important in this 
analysis, but their asymptotics in the region $x\sim \frac{1}{2}\Delta \gg 1$ is crucial. 
This asymptotics can be obtained by matching $[r(x), R(x)]$'s expressions (\ref{rRperturb})
with their asymptotics (\ref{rRlimit}). This is the key idea of the method. 
This matching gives the leading-order asymptotics of $(\tilde{r}, \tilde{R})$ as 
\begin{equation} \label{tails}
\left[\begin{array}{c} \tilde{r}(x) \\ \tilde{R}(x) \end{array}\right] 
\longrightarrow \left[\begin{array}{c} s_1 r_0(x-\Delta) \\
s_2 R_0(x-\Delta) \end{array}\right]
\longrightarrow 
\left[\begin{array}{c} s_1 c\: e^{x-\Delta} \\ s_2 C e^{\Omega(x-\Delta)} \end{array}\right], 
\hspace{0.6cm} x \sim  \frac{1}{2}\Delta \gg 1.
\end{equation}
As $x\to -\infty$, $(\tilde{r}, \tilde{R}) \to 0$. 

When Eq. (\ref{rRperturb}) is substituted into (\ref{r}) and (\ref{R}), and higher-order
terms dropped, the linearized equations for perturbations $(\tilde{r}, \tilde{R})$
are found to be 
\begin{equation} \label{tilderR}
L\left[\begin{array}{c} \tilde{r} \\ \tilde{R} \end{array}\right]=0, 
\end{equation}
where the linearization operator $L$ is
\begin{equation}
L=\left[\begin{array}{cc} \partial_{xx}-1+3r_0^2(x)+\beta R_0^2(x) & 2r_0(x)R_0(x)\\
2r_0(x)R_0(x) & \partial_{xx}-\Omega^2+3R_0^2(x)+\beta r_0^2(x) \end{array}\right],
\end{equation}
which is self-adjoint. Eq. (\ref{tilderR}) has one localized homogeneous
solution $[r'_0(x), R'_0(x)]^{\mbox{\scriptsize T}}$ when $\beta \ne 0$ and 1. 
Here the prime is the derivative, and the superscript ``T'' is the transpose of a vector. 
Thus, in order for the solution $(\tilde{r}, \tilde{R})$ of Eq. (\ref{tilderR}) to exist, 
a solvability condition
\begin{equation}
\int_{-\infty}^{x_m} [r'_0(x), R'_0(x)] L \left[\begin{array}{c} \tilde{r} \\ \tilde{R} \end{array}\right] dx=0
\end{equation}
must be satisfied. Here $x_m \sim \frac{1}{2}\Delta$.  
This condition can be simplified through integration by parts as
\begin{equation}
\left[\tilde{r}'(x)r'_0(x)-\tilde{r}(x)r''_0(x)+ \tilde{R}'(x)R'_0(x)-\tilde{R}(x)R''_0(x)\right]_{-\infty}^{x_m}=0.
\end{equation}
Finally, substitution of the asymptotics (\ref{largex}) and (\ref{tails}) into the above condition gives 
\begin{equation} \label{spacing0}
e^{-(1-\Omega)\Delta}=-\frac{s_1s_2C^2\Omega^2}{c^2}.
\end{equation}
This equation readily shows that in order for the bound state to exist, we must have $s_1s_2=-1$. 
In that case, the spacing formula (\ref{spacing0}) becomes exactly the same as (\ref{spacing}).
Hence the above simplified tail-matching method reproduces the results from the more general 
analysis in \cite{YangStudies98}.   

The relative errors of the leading-order bound states
(\ref{rsum}), (\ref{Rsum}), and the spacing formula (\ref{spacing})
have been discussed in \cite{YangStudies98}, and these errors 
are $O(e^{-\Delta}, e^{-\Omega\Delta})$. 
For the bound states, we have 
\begin{equation} \label{rerror}
r(x)=\left[ r_0(x)\mp r_0(x-\Delta) \right]\left[1+O(e^{-\Delta}, e^{-\Omega\Delta})\right], 
\end{equation}
\begin{equation} \label{Rerror}
R(x)=\left[R_0(x)\pm R_0(x-\Delta)\right] \left[1+O(e^{-\Delta}, e^{-\Omega\Delta})\right]. 
\end{equation}
These error estimates can also be obtained as follows. 
Let us consider the type-I bound state. Write
\begin{equation} 
r(x)=r_0(x)- r_0(x-\Delta) +\hat{r}(x), 
\end{equation}
\begin{equation} 
R(x)=R_0(x)+ R_0(x-\Delta) +\hat{R}(x), 
\end{equation}
where $\hat{r}, \hat{R} \ll 1$. Substituting these equations into (\ref{r}) and (\ref{R}), 
we find that in the region $-\infty < x \ll \Delta$, 
\begin{equation}\label{LhatrR}
L\left[\begin{array}{c}\hat{r} \\ \hat{R} \end{array} \right]
=\left[\begin{array}{c} 
-\left[3r_0^2(x)+\beta R_0^2(x)\right]r_0(x-\Delta)+2\beta r_0(x)R_0(x)R_0(x-\Delta) \\
\left[3R_0^2(x)+\beta r_0^2(x)\right]R_0(x-\Delta)-2\beta r_0(x)R_0(x)r_0(x-\Delta), 
\end{array} \right].
\end{equation}
Here terms which are higher-order than the ones kept in (\ref{LhatrR}) have been dropped. 
A similar equation can be obtained in the region $0\ll x < \infty$. 
Inspection of these equations shows that for all values of $x$, 
$(\hat{r}, \hat{R})$ is exponentially small compared to the leading-order terms
in (\ref{rsum}) and (\ref{Rsum}), and
the relative errors are $O(e^{-\Delta}, e^{-\Omega\Delta})$ as
shown in (\ref{rerror}) and (\ref{Rerror}). 

The spacing formula (\ref{spacing}) together with the results in \cite{YangPhysica97}
reveals that, in order for $\Delta \gg 1$, we must have $0<\beta <1$. 
In addition, we must have either $C/c\ll 1$, $\Omega<1$, or $C/c\gg 1$, $\Omega>1$. 
In the former limit, $R_0(x) \ll r_0(x)$, and $\Omega \approx \Omega_c$;
while in the latter limit, $r_0(x)\ll R_0(x)$, and $\Omega \approx 1/\Omega_c$. 
In both limits, the bound states look like two wave-and-daughter-waves glued together. 
These two limits are actually equivalent through
a scaling of variables (the so-called reciprocal relation in \cite{YangPhysica97}). 
Thus in the rest of this article, we will just consider the former limit where
$\Omega \approx \Omega_c$. Fundamental solitons in this limit have been 
perturbatively determined in \cite{YangPhysica97}. Utilizing those results, 
we readily find that the asymptotic formula for the spacing $\Delta$ is
\begin{equation} \label{Deltalimit}
\Delta \longrightarrow \frac{1}{1-\Omega_c}\left\{-\ln(\Omega-\Omega_c)+K+o(1)\right\}, 
\hspace{0.4cm} \Omega \to \Omega_c, 
\end{equation}
where the constant $K$ is 
\[K=(3-2\Omega_c)\ln 2-2\ln \Omega_c+\ln\gamma, \]
and
\begin{equation} \label{gamma}
\gamma=\frac{(1-\Omega_c^{3})\int^{\infty}_{-\infty}\mbox{sech}^{4\Omega_c}xdx}
                  {2\Omega_c \int^{\infty}_{-\infty}\mbox{sech}^{2\Omega_c}xdx}. 
\end{equation}

\section{Linear-stability analysis of two-vector-soliton \\ bound states}
To study the linear stability of the above two-vector-soliton bound states, 
we perturb these states as
\begin{equation}
A=e^{it}\left\{r(x)+\psi_1(x)e^{i\lambda t}+\psi_2^*(x)e^{-i\lambda^* t}\right\}, 
\end{equation}
\begin{equation}
B=e^{i\Omega^2 t}\left\{R(x)+\psi_3(x)e^{i\lambda t}+\psi_4^*(x)e^{-i\lambda^* t}\right\}, 
\end{equation}
where $[r(x), R(x)]$ is a two-vector-soliton bound state, 
$\psi_k\; (1\le k\le 4)$ are infinitesimal disturbances, 
$\lambda$ is the eigenvalue, and $\psi^*$ is the
complex conjugate of $\psi$. Substituting these equations into (\ref{A}) and (\ref{B})
and dropping higher-order terms, we get the following eigenvalue relation
\begin{equation} \label{LPsi}
{\cal L}\Psi=\lambda \Psi, 
\end{equation}
where the linearization operator ${\cal L}$ is
\tiny
\begin{equation} \label{L}
{\cal L}=\left( \begin{array}{cccc}
\partial_{xx}-1+2r^{2}+\beta R^{2} & r^{2} & \beta rR & \beta rR \\
-r^2 & -\partial_{xx}+1-2r^{2}-\beta R^{2} & -\beta rR & -\beta rR \\
\beta rR & \beta rR & \partial_{xx}-\Omega^2+2R^{2}+\beta r^{2} & R^{2} \\
-\beta rR & -\beta rR & -R^2 & -\partial_{xx}+\Omega^2-2R^{2}-\beta r^{2}
\end{array} \right), 
\end{equation}
\normalsize
and $\Psi=(\psi_1, \psi_2, \psi_3, \psi_4)^{\mbox{\scriptsize T}}$. 
The spectrum of ${\cal L}$ contains all information
on the linear stability of the two-vector-soliton bound state. 
If ${\cal L}$ has eigenvalues with Im($\lambda)<0$, then the bound state is linearly
unstable; otherwise, it is linearly stable. Obviously, 
the continuous spectrum of ${\cal L}$ always lies on the real axis, thus 
we only need to look at the discrete eigenvalues of ${\cal L}$. 
Notice that ${\cal L}^2$ is self-adjoint, thus
${\cal L}$'s eigenvalues are either purely real, or purely imaginary. 
In addition, if $\lambda$ is an eigenvalue of ${\cal L}$, so are  $-\lambda$.
Hence ${\cal L}$'s eigenvalues always come in pairs on the real
or imaginary axis. 

It is insightful to view ${\cal L}$'s discrete eigenvalues 
for the bound states $(r, R)$ in the perspective of 
eigenvalue bifurcations from a similar operator ${\cal L}_0$ 
for fundamental vector solitons $(r_0, R_0)$. Here 
\begin{equation} \label{L0}
{\cal L}_0={\cal L}|_{(r=r_0, R=R_0)}.
\end{equation}
The spectrum structure of ${\cal L}_0$ has been determined completely
in \cite{YangStudies97,PeliYang00}. 
This operator has 6 or 8 discrete eigenvalues (multiplicity counted), depending on whether
an internal mode exists or not. 
The zero eigenvalue always has multiplicity 6 ---
three from position and phase invariances (Goldstein modes), 
and the other three from velocity and 
frequency (or equivalently, amplitude) variations. 
When an internal mode exists, a pair of real eigenvalues of opposite sign
are present as well.
If two vector solitons form a widely-separated stationary bound state $(\Delta \gg 1)$, 
both solitons must be either wave-and-daughter-waves (with $\Omega\approx \Omega_c$ and $0<\beta<1$), 
or having nearly equal amplitudes in the two components (with $\Omega \approx 1$ and $\beta >0$) 
\cite{YangPhysica97,Yangunpublished}
(only the former type of bound states is studied in this paper). 
It has been shown in \cite{PeliYang00} that 
wave-and-daughter-waves do not admit internal modes. 
For instance, single vector solitons in Fig. 1's bound states 
do not have internal modes \cite{PeliYang00}. 
Thus ${\cal L}_0$ has only a single discrete eigenvalue zero of multiplicity 6. 
In a bound state of two wave-and-daughter-waves, 
${\cal L}$ will then have 12 discrete eigenvalues (multiplicities counted) ---
double that for a single wave-and-daughter-wave.
Here, no new discrete eigenvalues of ${\cal L}$ 
can be generated from edge bifurcations of ${\cal L}_0$
because the edge points of ${\cal L}_0$'s continuous spectrum 
are not in the continuous spectrum. 
Now the zero eigenvalue of ${\cal L}$ 
still has multiplicity 6. Another six eigenvalues of ${\cal L}$ must bifurcate from 
zero due to tail interactions between solitons. 
Tracking of these six bifurcated eigenvalues will then provide a complete
characterization of linear stability for these two-soliton bound states. 

In this section, we propose a new and general tail-matching method to derive
explicit analytical formulas for the six eigenvalues of ${\cal L}$ 
that bifurcate from zero, 
under the condition that the individual vector solitons in the bound state are widely separated.  
The idea of this method is to perturbatively determine the bifurcated eigenstate around each 
vector soliton. By matching their tail asymptotics in the center region of the eigenstate, 
their asymptotics in that region can then be explicitly obtained. 
Finally by utilizing the solvability conditions
for the eigenstate, analytical formulas for the bifurcated eigenvalues 
will be derived. These formulas turn out to be exactly the same as
those obtained by the KSGO method \cite{YangPRE01}, 
but the present method is much simpler. 

The bifurcated eigenvalues are of two types. 
One type consists of a pair of eigenvalues which bifurcate
from the position-related zero eigenvalue. At infinite soliton separation, 
these eigenstates are equal to the sum of two position-induced Goldstein modes 
separated infinitely apart. The other type consists of two pairs of eigenvalues 
which bifurcate from the phase-related zero eigenvalue. At infinite soliton separation, 
these eigenstates are equal to the sum of two phase-induced Goldstein modes 
separated infinitely apart.
These two types of eigenvalues have their counterparts in 
the linear stability analysis by the KSGO method \cite{YangPRE01}. 

Below, we consider eigenvalue bifurcations in type-I and II bound states.  
It turns out that eigenvalues of type-II bound states
are simply equal to those of type-I states multiplied by $i$ (see also \cite{YangPRE01}). 
Thus calculations for only type-I states will be presented. 
These states are anti-symmetric in $r$, and symmetric in $R$, around the center of the bound states
[see Fig. 1(a)]. When $\Delta \to \infty$, 
\begin{equation} \label{typeI}
r(x)\longrightarrow r_0(x)- r_0(x-\Delta), \hspace{0.5cm} 
R(x)\longrightarrow R_0(x)+ R_0(x-\Delta). 
\end{equation}
Bifurcations of position- and phase-related eigenvalues are studied separately next.  

\subsection{Bifurcation of position-related eigenvalues}
When $\Delta \to \infty$, the eigenstate bifurcated from the position-related zero eigenvalue 
is a sum of two position-induced Goldstein modes of fundamental vector solitons: 
\begin{equation} \label{Psilimit}
\Psi(x) \longrightarrow \Psi_0(x)+\hat{\Psi}_0(x), \hspace{0.5cm} \Delta \to \infty, 
\end{equation}
where 
\begin{equation}\label{Psi0po}
\Psi_0(x)=\left[\begin{array}{c} r'_0(x) \\ r'_0(x) \\ R'_0(x) \\ R'_0(x)\end{array}\right], \hspace{0.5cm}
\hat{\Psi}_0(x)=\left[\begin{array}{c} r'_0(x-\Delta) \\ r'_0(x-\Delta) \\ -R'_0(x-\Delta) \\ -R'_0(x-\Delta)
\end{array}\right].  
\end{equation}
The first two components of $\Psi$ are anti-symmetric around $x=\frac{1}{2}\Delta$,  
and the last two components are symmetric around $x=\frac{1}{2}\Delta$.   
Note that in the same limit, the eigenstate $\Psi_0(x)-\hat{\Psi}_0(x)$ is 
simply $[r'(x),\; r'(x), \; R'(x), \; R'(x)]^{\mbox{\scriptsize T}}$, 
which is the un-bifurcated Goldstein eigenmode with eigenvalue zero and is thus not considered.  

In the limit $\Delta \gg 1$, 
we consider the bifurcated eigenstate in the region $-\infty < x \ll \Delta$, and
expand it as a perturbation series in powers of the small eigenvalue $\lambda$:
\begin{equation}\label{Psiexpansion}
\Psi(x)=\Psi_0(x)+\lambda \Psi_1(x)+\lambda^2\Psi_2(x) +o(\lambda^2), 
\hspace{0.8cm} -\infty < x \ll \Delta.  
\end{equation}
In this region, we also expand
\begin{equation} \label{Lexpansion}
{\cal L}={\cal L}_0+\lambda^2 {\cal L}_2+o(\lambda^2). 
\end{equation}
It is noted that when $\Delta \gg 1$, single solitons in the bound state considered
are wave-and-daughter-waves whose two components have different orders
(one component asymptotically much smaller than the other). 
This fact certainly has implications in the stability analysis. 
In particular, different components of 
$\Psi$ and ${\cal L}$ in the region $-\infty < x \ll \Delta$
may have slightly different orders of magnitude. 
Thus, a perturbation expansion with a more-detailed ordering of 
different components than that in (\ref{Psiexpansion}) and (\ref{Lexpansion})
might be needed. But as we will see next, 
(\ref{Psiexpansion}) and (\ref{Lexpansion}) still work. 


Now we substitute the perturbation expansions (\ref{Psiexpansion}) and
(\ref{Lexpansion}) into the eigenvalue
equation (\ref{LPsi}). At $O(1)$, we get ${\cal L}_0\Psi_0(x)=0$ which is satisfied automatically. 
At $O(\lambda)$, we get
\begin{equation} \label{Psi1}
{\cal L}_0\Psi_1=\Psi_0, 
\end{equation}
whose solution is 
\begin{equation} \label{Psi1solution}
\Psi_1(x)=[\frac{1}{2}xr_0(x), \;\; -\frac{1}{2}xr_0(x), \;\;
 \frac{1}{2}xR_0(x), \;\; -\frac{1}{2}xR_0(x)]^{\mbox{\scriptsize T}}.
\end{equation}
Note that this function is an inhomogeneous solution of Eq. (\ref{Psi1}). 
In general, $\Psi_1$ should also include the homogeneous solutions
which are a linear combination of the three Goldstein modes: $\Psi_0(x)$ in 
(\ref{Psi0po}), and [$\Phi_0^{(1)}(x)$, $\Phi_0^{(2)}(x)$] in (\ref{Phi012}). 
But the $\Psi_0(x)$ term in $\Psi_1$ can be combined with the $O(1)$ term in
the expansion (\ref{Psiexpansion}), and the other two Goldstein modes 
[$\Phi_0^{(1)}(x)$, $\Phi_0^{(2)}(x)$] 
are phase-related and do not arise here. Thus the solution of Eq. (\ref{Psi1}) can 
be taken as (\ref{Psi1solution}) without loss of generality. 

At $O(\lambda^2)$, we get 
\begin{equation} \label{Psi2}
{\cal L}_0\Psi_2=\Psi_1-{\cal L}_2\Psi_0. 
\end{equation}
It is more convenient to express ${\cal L}_2\Psi_0$ in a different form. 
Recall that the un-bifurcated position-related Goldstein mode $\Psi_g(x)$ of ${\cal L}$ has 
the leading-order asymptotics $\Psi_0(x)-\hat{\Psi}_0(x)$ for $\Delta \gg 1$. 
When this asymptotics and ${\cal L}$'s expansion (\ref{Lexpansion})
are substituted into the Goldstein-mode relation 
${\cal L}\Psi_g(x)=0$, we find that asymptotically, 
\begin{equation}
\lambda^2 {\cal L}_2\Psi_0={\cal L}_0\hat{\Psi}_0.
\end{equation}
Thus Eq. (\ref{Psi2}) can be rewritten as
\begin{equation} \label{Psi2new}
{\cal L}_0\left[\Psi_2+\lambda^{-2}\hat{\Psi}_0\right]
=\Psi_1. 
\end{equation}
Note that ${\cal L}_0\hat{\Psi}_0$ is $O(\lambda^2)$, thus
the above equation is not dis-ordered. 

The solvability condition of Eq. (\ref{Psi2new})
will produce formulas for the eigenvalue $\lambda$. 
To do this, we need the asymptotics of function $\Psi_2$ in the region $x\sim \frac{1}{2}\Delta \gg 1$, 
which we derive using the tail-matching idea. 
We have known that $\Psi$'s leading-order asymptotics
at $\Delta \gg 1$ is given by (\ref{Psilimit}) for all values of $x$. 
Combining this asymptotics with the perturbation expansion (\ref{Psiexpansion}) and
solutions (\ref{Psi0po}) and (\ref{Psi1solution}), we see that  
\begin{equation}\label{Psi2limit}
\Psi_2(x) \to \frac{1}{\lambda^2}
\hat{\Psi}_0(x), 
\hspace{0.5cm} x \sim \frac{1}{2}\Delta \gg 1.
\end{equation}
Of course, $\Psi_2(x) \to 0$ when $x \to -\infty$. 

The homogeneous equation of (\ref{Psi2new}) has three linearly independent solutions which 
are the Goldstein modes $\Psi_0(x)$ in 
(\ref{Psi0po}) and [$\Phi_0^{(1)}(x), \Phi_0^{(2)}(x)$] in (\ref{Phi012}). 
Thus Eq. (\ref{Psi2new}) has three solvability conditions. 
These conditions can be readily derived by noting that 
diag$(1,-1,1,-1){\cal L}_0$ is self-adjoint. 
It turns out that two of the
solvability conditions induced by the Goldstein modes [$\Phi_0^{(1)}, \Phi_0^{(2)}$]
are satisfied automatically. The remaining solvability condition reads, 
\begin{eqnarray}
\int_{-\infty}^{x_m} \langle\Psi_0 | \mbox{diag} (1,-1,1,-1) | {\cal L}_0
(\Psi_2 +\lambda^{-2}\hat{\Psi}_0)\rangle dx  \nonumber \\
=\int_{-\infty}^{x_m} \langle\Psi_0 | \mbox{diag} (1,-1,1,-1) | \Psi_1 \rangle dx,
\end{eqnarray}
where $x_m \sim \frac{1}{2}\Delta$, 
$\langle \cdot |$ and $| \cdot \rangle$ are the Dirac ket and bra notations \cite{dirac}. 
Integrating by parts to its left-hand-side and
simplifying its right-hand-side, we get
\begin{equation} \label{temp}
\left[\langle\Psi_0 | (\Psi_{2x}+\lambda^{-2}\hat{\Psi}_{0x})
\rangle-\langle\Psi_{0x} | 
(\Psi_{2}+\lambda^{-2}\hat{\Psi}_0)
\rangle\right]_{-\infty}^{x_m}
=-\frac{1}{2}\int_{-\infty}^{x_m}(r_0^2+R_0^2)dx. 
\end{equation}
The left-hand-side of (\ref{temp}) can be calculated using the asymptotics 
(\ref{largex}), (\ref{Psi2limit}) and equation (\ref{Psi0po}), 
while the integral on the 
right-hand-side of (\ref{temp}) is asymptotically equal to a similar integral but with the upper limit $x_m$
replaced by $\infty$. After these calculations, the eigenvalue $\lambda$ is finally found to be
\begin{equation} \label{positionlambda}
\lambda^2=\frac{16(1-\Omega)c^2e^{-\Delta}}{M+N}, 
\end{equation}
where $M$ and $N$ are the masses of the $r$ and $R$ components in a fundamental vector soliton: 
\begin{equation}
M\equiv \int_{-\infty}^\infty r_0^2 dx,  \hspace{0.5cm}
N\equiv \int_{-\infty}^\infty R_0^2 dx.
\end{equation}
We immediately see that formula (\ref{positionlambda}) is 
identical to the one derived in \cite{YangPRE01} using
the KSGO method. 
Thus, the present tail-matching method has the same accuracy as 
the KSGO method, but is only simpler. 
Recall that we only consider the $\Omega \approx \Omega_c (<1)$ limit. 
In this limit, the type-I state flips sign in its larger component, 
and does not flip sign in its smaller component [see Fig. 1(a)]. 
According to formula (\ref{positionlambda}), this position-related eigenvalue
$\lambda$ is real, thus it does not create instability. 
A comparison between the analytical formula (\ref{positionlambda})
and numerical values at $\beta=\frac{2}{3}$
has been made in \cite{YangPRE01}, and excellent agreement has been obtained. 

Formula (\ref{positionlambda}) shows that $\lambda=O(e^{-\frac{1}{2}\Delta})$. 
When $\Omega \to \Omega_c$ and $0<\beta<1$, $\Delta \to \infty$ (see Sec. 2). 
In this limit, the asymptotic formula for $\lambda$ can be obtained more explicitly
from (\ref{Deltalimit}) and (\ref{positionlambda}) as 
\begin{equation}
\lambda^2 \longrightarrow  \alpha \: (\Omega-\Omega_c)^{\frac{1}{1-\Omega_c}}, 
\hspace{0.5cm} \Omega \to \Omega_c, 
\end{equation}
where the constant $\alpha$ is
\begin{equation}
\alpha=32(1-\Omega_c)\left(\frac{\Omega_c^2 4^{\Omega_c}}{8\gamma}\right)^{\frac{1}{1-\Omega_c}}.
\end{equation}
Here $\Omega_c$ and $\gamma$ are defined in Eqs. (\ref{Omegac}) and (\ref{gamma}). 

\subsection{Bifurcation of phase-related eigenvalues}
Calculations for the bifurcation of phase-related eigenvalues are quite 
similar to that done above. In this case, the eigenmode has the following asymptotics: 
\begin{equation}\label{Psilimit2}
\Psi(x) \longrightarrow \Phi_0(x)+\hat{\Phi}_0(x), \hspace{0.5cm} \Delta \to \infty, 
\end{equation}
where 
\begin{equation} \label{Phi0ph}
\Phi_0(x)=\Phi_{0}^{(1)}(x)+\delta \Phi_0^{(2)}(x), 
\end{equation}
\begin{equation}
\hat{\Phi}_0(x)=\Phi_{0}^{(1)}(x-\Delta)-\delta \Phi_0^{(2)}(x-\Delta), 
\end{equation}
\begin{equation}\label{Phi012}
\Phi_0^{(1)}(x)=\left[\begin{array}{c} r_0(x) \\ -r_0(x) \\ 0 \\ 0\end{array} \right], \hspace{0.5cm}
\Phi_0^{(2)}(x)=\left[\begin{array}{c} 0 \\ 0 \\ R_0(x) \\ -R_0(x) \end{array} \right],
\end{equation}
and $\delta$ is some constant (to be determined). 
The first two components of this mode are symmetric around $x=\frac{1}{2}\Delta$,  
and the last two components are anti-symmetric around $x=\frac{1}{2}\Delta$.   
Note that the eigenstate 
with asymptotics $\Phi_0(x)-\hat{\Phi}_0(x)$ is 
simply $[r(x),\; -r(x), \; \delta R(x), \; -\delta R(x)]^{\mbox{\scriptsize T}}$, 
which is the un-bifurcated Goldstein eigenmode with eigenvalue zero and is not a concern.  

Next, we construct a perturbation-series solution for $\Psi(x)$ in the region 
$-\infty<x\ll \Delta$. The perturbation series is similar to (\ref{Psiexpansion}): 
\begin{equation}\label{Psiexpansion2}
\Psi(x)=\Phi_0(x)+\lambda \Phi_1(x)+\lambda^2\Phi_2(x) +o(\lambda^2), \hspace{0.8cm} -\infty < x \ll \Delta,  
\end{equation}
where $\Phi_0(x)$ is given by (\ref{Phi0ph}). Substituting this expansion 
and (\ref{Lexpansion}) into the eigenvalue relation (\ref{LPsi}), 
the $O(1)$ equation is satisfied automatically. At $O(\lambda)$, 
we get 
\begin{equation} \label{Psi1b}
{\cal L}_0\Phi_1=\Phi_0, 
\end{equation}
whose solution is 
\begin{equation} \label{Psi1solution2}
\Phi_1(x)=\frac{1}{2} \frac{\partial}{\partial \omega}
\left[\begin{array}{c} r_0(x; \omega, \Omega) \\ r_0(x; \omega, \Omega) 
\\ R_0(x; \omega, \Omega) \\ R_0(x; \omega, \Omega) \end{array}\right]_{\omega=1}
+\frac{\delta}{2\Omega} \frac{\partial}{\partial \Omega}
\left[\begin{array}{c} r_0(x; \omega, \Omega) \\ r_0(x; \omega, \Omega) 
\\ R_0(x; \omega, \Omega) \\ R_0(x; \omega, \Omega) \end{array}\right]_{\omega=1}. 
\end{equation}
Here $(r_0, R_0)$ is the fundamental vector soliton obtained from ODEs (\ref{r}) and (\ref{R})
without rescaling of $\omega=1$. 
Again, the homogeneous solution of Eq. (\ref{Psi1b}), which is a linear combination
of Goldstein modes $\Psi_0(x)$ in (\ref{Psi0po}) and [$\Phi_0^{(1)}(x), \Phi_0^{(2)}(x)$]
in (\ref{Phi012}), is not included because the latter modes can be absorbed into 
the $O(1)$ term in the perturbation expansion (\ref{Psiexpansion2}), and the former mode does not arise. 

At $O(\lambda^2)$, we get
\begin{equation} \label{Psi2b}
{\cal L}_0\Phi_2=\Phi_1-{\cal L}_2\Phi_0.  
\end{equation}
Again, utilizing the un-bifurcated phase-related Goldstein modes 
of ${\cal L}$ with asymptotics $\Phi_0(x)-\hat{\Phi}_0(x)$ ---
similar to what we have done in Sec. 3.1, 
we can rewrite ${\cal L}_2\Phi_0$ so that Eq. (\ref{Psi2b}) becomes
\begin{equation}
{\cal L}_0\left[\Phi_2+\lambda^{-2}\hat{\Phi}_0\right]
=\Phi_1. 
\end{equation} 
This equation has three solvability conditions induced by the three Goldstein modes in the homogeneous solution. 
The condition induced by the mode $\Psi_0(x)$ in (\ref{Psi0po}) is satisfied automatically. 
The other two conditions are 
\begin{eqnarray}
\int_{-\infty}^{x_m} \langle\Phi_0^{(k)} | \mbox{diag} (1,-1,1,-1) | 
{\cal L}_0(\Phi_2+\lambda^{-2}\hat{\Phi}_0) \rangle dx   \nonumber \\
=\int_{-\infty}^{x_m} \langle\Phi_0^{(k)} | \mbox{diag} (1,-1,1,-1) | \Phi_1 \rangle dx, 
\end{eqnarray}
where $x_m \sim \frac{1}{2}\Delta$, and $k=1, 2$. 
Integration by parts simplifies these conditions as
\begin{equation}\label{temp1}
\left[\langle\Phi_0^{(1)} | (\Phi_{2x}+\lambda^{-2}\hat{\Phi}_{0x})
\rangle-\langle\Phi_{0x}^{(1)} | (\Phi_{2}+\lambda^{-2}\hat{\Phi}_0)
\rangle\right]_{-\infty}^{x_m}
=\frac{1}{2}M_{\omega}+\frac{\delta}{2\Omega}M_{\Omega}, 
\end{equation}
and 
\begin{equation}\label{temp2}
\left[\langle\Phi_0^{(2)} | (\Phi_{2x}+\lambda^{-2}\hat{\Phi}_{0x})
\rangle-\langle\Phi_{0x}^{(2)} | 
(\Phi_{2}+\lambda^{-2}\hat{\Phi}_0)
\rangle\right]_{-\infty}^{x_m}
=\frac{1}{2}N_{\omega}+\frac{\delta}{2\Omega}N_{\Omega}.  
\end{equation}
To calculate the left-hand-sides of the above two equations, we need 
the asymptotics of function $\Phi_2(x)$ in the region $x \sim \frac{1}{2}\Delta$. 
This asymptotics can be obtained by comparing $\Psi(x)$'s asymptotics (\ref{Psilimit2})
with its perturbation expansion (\ref{Psiexpansion2}). 
This comparison shows that $\Phi_2(x)$ must have the asymptotics
\begin{equation}
\Phi_2(x) \to \frac{1}{\lambda^2}\hat{\Phi}_0(x), 
\hspace{0.5cm} x \sim \frac{1}{2}\Delta \gg 1.
\end{equation}
When this asymptotics as well as (\ref{largex}) is substituted 
into Eqs. (\ref{temp1}) and (\ref{temp2}) and parameter $\delta$ eliminated, 
the eigenvalue $\lambda$ is then given by the quartic equation
\begin{equation}\label{quartic}
\lambda^4-\frac{16c^2(N_\Omega-M_\omega)e^{-\Delta}}{M_\omega N_\Omega-M_\Omega N_\omega}\lambda^2
-\frac{16^2c^4e^{-2\Delta}}{M_\omega N_\Omega-M_\Omega N_\omega}=0. 
\end{equation}
Again, this formula for phase-related eigenvalues is identical to that obtained 
in \cite{YangPRE01} by the KSGO method. 
As pointed in \cite{YangPRE01}, this formula shows that a two-vector-soliton bound state
always has one phase-related unstable eigenvalue which bifurcates from zero, thus is always 
linearly unstable. Comparison between this formula and numerical values has also been made in 
\cite{YangPRE01}, and excellent agreement has been observed.

Formula (\ref{quartic}) shows that phase-related eigenvalues $\lambda=O(e^{-\frac{1}{2}\Delta})$, 
the same as position-related eigenvalues [see Eq. (\ref{positionlambda})]. 
The asymptotics of these eigenvalues in the limit $\Omega \to \Omega_c$ can also be
obtained from (\ref{Deltalimit}) and (\ref{quartic}), but is not pursued here. 

Next, we briefly discuss the linear stability of type-II vector-soliton bound states [see Fig. 1(b)]. 
In the limit of 
large separation, these solitons have the asymptotics 
\begin{equation} \label{typeII}
r(x)\longrightarrow r_0(x)+ r_0(x-\Delta), \hspace{0.5cm} 
R(x)\longrightarrow R_0(x)- R_0(x-\Delta). 
\end{equation}
Repeating the above analytical calculations, we find that
the eigenvalues for type-II states are equal to those of type-I states multiplied by $i$.
Thus type-II states are always linearly unstable as well.
But different from type-I states, the instability of type-II states 
in the limit $\Omega \approx \Omega_c (<1)$ 
is caused by two unstable eigenvalues: one related to position-mode bifurcations
[see (\ref{positionlambda})], and the other one related to phase-mode bifurcations
[see (\ref{quartic})]. 
This result agrees with that by 
the KSGO method as well as the numerical result \cite{YangPRE01}.

\section{Discussion}

In this paper, we have analytically studied the linear stability of two-vector-soliton
bound states in the coupled NLS equations by a new tail-matching method. 
Under the condition that the two vector solitons are wave-and-daughter-waves and
are widely separated, we have
calculated small eigenvalues of these bound states 
that bifurcate from the zero eigenvalue of single vector solitons. 
These small eigenvalues calculated are all the discrete non-zero eigenvalues of 
the bound states. 
We have shown that these bound states are always linearly unstable due to 
the existence of one unstable phase-induced eigenvalue. 
The analytical formulas for eigenvalues derived from this tail-matching method
turn out to be exactly the same as those from the KSGO method \cite{YangPRE01}, but
the present method is much simpler. 
Even though our calculations were performed for two-vector-soliton 
bound states, these calculations can apparently be generalized to 
$n$-vector-soliton bound states.

This tail-matching method for the linear stability analysis of 
multi-soliton bound states is apparently quite general, and it can be applied
to a wide range of other wave systems where similar multi-pulse
bound states have been reported \cite{YangStudies98,ChampneysToland,ChampneysToland2}. 
In addition, this method should be applicable to the stability analysis of
other types of multi-pulses from nonlocal bifurcations in coupled-NLS-type equations
\cite{ChampneysYang}. Preliminary linear stability results through numerical studies
shows that such multi-pulses are also linearly unstable \cite{ChampneysYang}, consistent with the 
results of the present article. 
This tail-matching method 
can also be used to calculate eigenvalue bifurcations of multi-pulses
from internal modes (non-zero eigenvalues) of single pulses --- 
a bifurcation which incidentally does not arise in the present problem 
because internal modes do not exist in single wave-and-daughter-waves
\cite{PeliYang00}. 

Under what conditions can this tail-matching method give useful results? 
The main condition is that the individual pulses in the
multi-pulse bound state are widely separated. 
This condition will dictate in what parameter regions 
such multi-pulses can exist, and thus tail-matching can proceed.
For instance, for the coupled NLS equations (\ref{A}) and (\ref{B}), 
the spacing formulas (\ref{spacing}) and (\ref{Deltalimit}) 
dictate that widely-separated multi-vector-soliton bound states
($\Delta \gg 1$) exist when $\Omega$ is near the local bifurcation
boundary $\Omega=\Omega_c$. This is precisely the parameter region
where our tail-matching linear stability analysis is performed.

The two-vector-soliton bound states studied in this paper reside outside
the continuous spectrum of the corresponding linear-wave system.
It is known that multi-pulse embedded solitons residing inside the
continuous spectrum exist in various wave systems as well 
\cite{akylaskung,klauder,calvoakylas,champneys03,yangakylas}. 
An interesting open issue is whether the tail-matching method 
can also be applied to the linear stability analysis of such
multi-pulse embedded solitons. In the third-order NLS equation, 
it has been shown numerically that such embedded solitons 
are all linearly stable, but nonlinearly semi-stable \cite{yangakylas}.

Lastly, we relate this tail-matching method to other existing 
techniques for the linear stability analysis of multi-pulse bound states. 
Currently, the following techniques exist: 
the KSGO method \cite{YangPRE01}, 
the dynamical-system method \cite{Yew97,Sandstede98}, and 
the effective-interaction-potential method \cite{malomed,malomed2,BuryakAkhmediev95}. 
The present method is asymptotically accurate. It gives 
the same results as the KSGO method, but is much simpler. 
The dynamical-system method can count the number of
unstable eigenvalues, or express the eigenvalues as the zeros of the
Evans function. But it generally does not produce explicit formulas for eigenvalues. 
The effective-potential method can only capture position-related eigenvalue
bifurcations, not phase-related eigenvalue bifurcations.
(For the coupled NLS equations, phase-related eigenvalues are more important). 
In view of this comparison, we feel that the
tail-matching method for the linear stability of multi-soliton bound states
is very promising.

\begin{center}
Department of Mathematics and Statistics \\
University of Vermont \\
Burlington, VT 05401, USA \\
\end{center}
\end{document}